\documentclass[final,5p,times,twocolumn]{elsarticle}

\usepackage{graphicx,times,amsmath,amssymb,mathrsfs,subfigure,stfloats,booktabs,url,multirow}
\usepackage[lined,ruled,commentsnumbered,linesnumbered]{algorithm2e}

\begin{document}

\begin{frontmatter}

\title{CSP-Net: Common Spatial Pattern Empowered Neural Networks for EEG-Based Motor Imagery Classification}

\author[AIA,SZ]{Xue~Jiang}
\author[AIA,SZ]{Lubin~Meng}
\author[AIA,SZ]{Xinru~Chen}
\author[AIA,SZ]{Yifan~Xu}
\author[AIA,SZ]{Dongrui~Wu\corref{cor1}}
\cortext[cor1]{Emails: xuejiang@hust.edu.cn (Xue Jiang), lubinmeng@hust.edu.cn (Lubin Meng), xrchen@hust.edu.cn (Xinru Chen), yfxu@hust.edu.cn (Yifan Xu), drwu@hust.edu.cn (Dongrui Wu).
Dongrui Wu is the corresponding author.}

\address[AIA]{Key Laboratory of the Ministry of Education for Image Processing and Intelligent Control, School of Artificial Intelligence and Automation, Huazhong University of Science and Technology, Wuhan 430074, China}
\address[SZ]{Shenzhen Huazhong University of Science and Technology Research Institute, Shenzhen 518063, China}

\begin{abstract}
Electroencephalogram-based motor imagery (MI) classification is an important paradigm of non-invasive brain-computer interfaces. Common spatial pattern (CSP), which exploits different energy distributions on the scalp while performing different MI tasks, is very popular in MI classification. Convolutional neural networks (CNNs) have also achieved great success, due to their powerful learning capabilities. 
This paper proposes two CSP-empowered neural networks (CSP-Nets), which integrate knowledge-driven CSP filters with data-driven CNNs to enhance the performance in MI classification. CSP-Net-1 directly adds a CSP layer before a CNN to improve the input discriminability. CSP-Net-2 replaces a convolutional layer in CNN with a CSP layer. The CSP layer parameters in both CSP-Nets are initialized with CSP filters designed from the training data. During training, they can either be kept fixed or optimized using gradient descent. Experiments on four public MI datasets demonstrated that the two CSP-Nets consistently improved over their CNN backbones, in both within-subject and cross-subject classifications. They are particularly useful when the number of training samples is very small. Our work demonstrates the advantage of integrating knowledge-driven traditional machine learning with data-driven deep learning in EEG-based brain-computer interfaces.
\end{abstract}


\end{frontmatter}

\section{Introduction}

A brain-computer interface (BCI) establishes a direct communication pathway that enables the human brain to interact with external devices \cite{BCIIntro}. Electroencephalogram (EEG), which records the electrical activities on the scalp of the brain, is the most widely used input signal in non-invasive BCIs due to its affordability and convenience \cite{BCIReview}. EEG-based BCIs have been used in controlling robots \cite{hochberg2012}, decoding speech \cite{anumanchipalli2019}, enhancing computer gaming experience \cite{krauledat2008}, and so on.

Motor imagery (MI) \cite{MI2001} is a classical paradigm of EEG-based BCIs, where a subject imagines the movement of a body part, e.g., right hand, left hand, right foot, left foot, both feet, and/or tongue, without actually executing it. An MI induces changes in the sensory-motor rhythms (SMR) of corresponding areas of the cerebral cortex, which primarily involve modulations of the $\mu$ rhythm (8-12Hz) and the $\beta$ rhythm (14-30Hz) \cite{jeannerod1995}. Specifically, when an MI starts, these rhythmic activities decrease, resulting in event-related desynchronization (ERD); at the end of an MI, these rhythmic activities increase, resulting in event-related synchronization (ERS) \cite{pfurtscheller1997,blankertz2010}. Therefore, the detection of SMR patterns within specific areas of the cerebral cortex can be used to identify which body part the subject is imagining moving.

Many algorithms have been proposed for EEG-based MI classification. Common spatial pattern (CSP) \cite{ramoser2000,Blankertz2008} is one of the most widely used and effective approaches, which converts the raw multi-channel EEG signals into more discriminative spatial patterns. It was initially proposed for binary classification, by designing spatial filters that maximize the variance ratio of the filtered signals of different classes \cite{ramoser2000}. Dornhege \emph{et al.} \cite{dornhege2004} extended it to multi-class classification using a one-versus-the-rest strategy. Ang \emph{et al.} \cite{FBCSP} proposed filter bank CSP (FBCSP), which bandpass filters EEG signals into multiple frequency bands, extracts CSP features from each band, and then selects the most useful features for classification. Lotte \emph{et al.} \cite{lotte2010} introduced regularized CSP to enhance the robustness of CSP.

Recent years have witnessed significant increase in using deep learning for EEG signal decoding \cite{craik2019}, which integrates feature extraction and classification into a single end-to-end network. Among various deep architectures, convolutional neural networks (CNNs) are the most prevalent for MI classification~\cite{Altaheri2023, AlSaegh2021}. For example, Schirrmeister \emph{et al.} \cite{MNE} proposed ShallowCNN and DeepCNN for raw EEG classification. ShallowCNN is inspired by FBCSP and includes components such as temporal convolution, spatial convolution, log-variance calculation and a classifier, each corresponding to a specific step in FBCSP. DeepCNN is similar but includes more convolutional and pooling layers. Lawhern \emph{et al.} \cite{EEGNet} introduced a compact EEGNet, which has demonstrated promising performance across various BCI tasks, including MI classification. Inspired also by FBCSP, EEGNet uses a two-step sequence of temporal convolution followed by depthwise convolution. Recently, FBCNet~\cite{Mane2021} extends the FBCSP approach by utilizing a hierarchical architecture that enhances feature extraction through multi-dimensional filtering, allowing it to capture richer spatial and temporal patterns in EEG data. EEGConformer~\cite{Song2023} adopts a transformer-like architecture, combining self-attention mechanisms with convolutional layers, which enables the model to learn long-range dependencies in EEG signals effectively.

Though these data-driven deep models have achieved promising performance in MI classification, they usually require a large amount of labeled training data, which may not be always available in practice. This highlights the need to incorporate prior knowledge into EEG networks, as it can help reduce the reliance on extensive labeled datasets. By integrating prior knowledge, models can leverage existing insights about EEG signal characteristics, enhancing their generalization capabilities and performance even in data-scarce environments. This paper proposes CSP empowered neural networks (CSP-Net), which more effectively integrate CSP and CNNs. More specifically, we propose two CSP-Nets, by embedding CSP into different layers of the CNN models. The first, CSP-Net-1, places a CSP layer before a CNN to filter the EEG signals for enhancing their discriminability. The second, CSP-Net-2, replaces a CNN's convolutional layer with a CSP layer to provide task-specific prior knowledge initialization. The parameters in the CSP layer of both CSP-Nets are initialized from CSP filters designed on the training data. They can be fixed or optimized by gradient descent during training. In summary, CSP-Nets integrate the strengths of traditional CSP feature extraction with deep learning by embedding CSP layers in CNN architectures. This approach enhances the model's ability to capture relevant features from EEG signals, making it a more effective solution for MI classification. Our main contributions are:
\begin{itemize}
\item Integration of CSP and CNNs: We propose a novel framework that combines CSP with CNNs for MI classification, enhancing EEG feature extraction and improving classification performance.
\item Two CSP-Net Variants: We introduce two architectures, CSP-Net-1, which incorporates a CSP layer before the CNN, and CSP-Net-2, which replaces a convolutional layer with a CSP layer for task-specific prior knowledge initialization. Both variants allow CSP layer parameters to be fixed or further optimized.
\item Performance on Multiple EEG Datasets: CSP-Nets demonstrate strong performance across various scenarios, including within-subject and cross-subject classifications, as well as in small sample settings. The models demonstrate generalization across different backbone architectures, validated on four public MI datasets.
\end{itemize}

The rest of this paper is structured as follows. Section~\ref{sect:method} introduces the classical CSP and proposes two CSP-Nets. Section~\ref{sect:experiment} presents the experimental settings and experimental results. Finally, Section~\ref{sect:conclusion} draws conclusions.

\section{Methods}\label{sect:method}

This section introduces the CSP algorithm, five CNN models for MI classification, and our proposed two CSP-Nets to integrate CSP and CNNs.

\subsection{CSP}

CSP was first proposed by Koles \emph{et al.} \cite{Koles1990} to extract discriminative features from EEG signals of two human populations. M{\"u}eller-Gerking \emph{et al.} \cite{MuellerGerking1999} later extended it to MI classification. Since then, it has become one of the most popular and effective algorithms in MI-based BCIs \cite{ramoser2000,Blankertz2008}.

\begin{figure}[htbp]\centering
\subfigure[]{\label{fig:csp_before}  \includegraphics[width=.45\linewidth,clip]{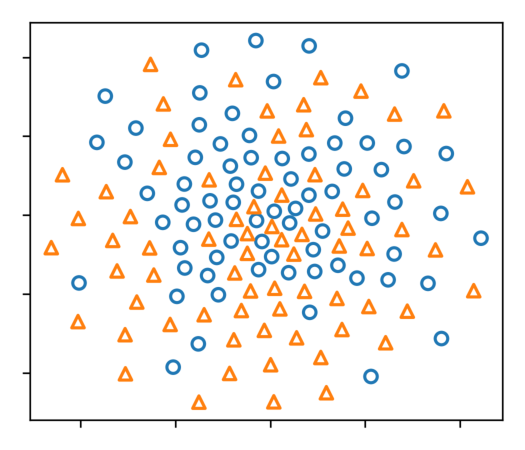}}
\subfigure[]{\label{fig:csp_after}   \includegraphics[width=.45\linewidth,clip]{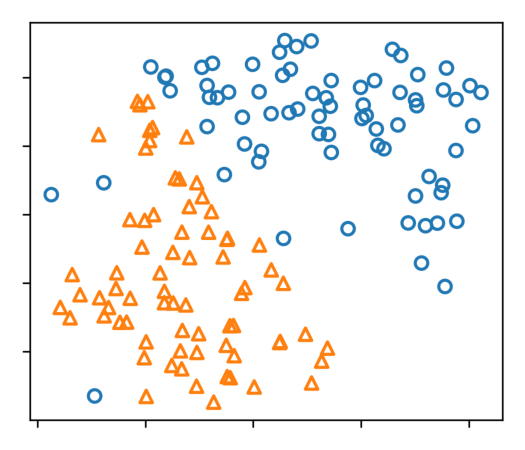}}
\caption{\emph{t}-SNE visualization of (a) the raw EEG trials; and, (b) the CSP-filtered trials. Different shapes (colors) represent different classes.} \label{fig:csp}
\end{figure}

Fig.~\ref{fig:csp} shows $t$-SNE visualization of some real EEG trials before and after CSP filtering from Subject~3 of Dataset 2a in BCI Competition IV \cite{MI4C}. Clearly, after CSP filtering, samples from different classes become more distinguishable.

For binary classification, CSP aims to learn spatial filters that maximize the variance of EEG signals from one class while simultaneously minimizing the variance from the other class. Let $X_i\in\mathbb{R}^{c\times t}$ be an EEG trial of MI task $i$, where $i\in\{1,2\}$ is the class index, $c$ the number of channels, and $t$ the number of time domain samples. CSP generates a spatial filtering matrix $W\in \mathbb{R}^{c\times f}$ ($f<c$) that projects the original EEG trials into a lower-dimensional space with higher discriminability. $W$ is obtained by maximizing (or minimizing):
\begin{align}
J(W)=\frac{W^{\top} \bar{X}_1 \bar{X}_1^{\top} W}{W^{\top} \bar{X}_2 \bar{X}_2^{\top} W}=\frac{W^{\top} \bar{C}_1 W}{W^{\top} \bar{C}_2 W}, \label{eq:csp_objective}
\end{align}
where $\bar{X}_i\in \mathbb{R}^{c\times t}$ is the averaged EEG trial from class $i$, and $\bar{C}_i\in \mathbb{R}^{c\times c}$ the mean spatial covariance matrix of all EEG trials in class $i$.

Since $J(W)=J(kW)$ for any arbitrary real constant $k$, maximizing $J(W)$ is equivalent to maximizing $W^{\top}\bar{C}_1 W$, subject to the constraint $W^{\top} \bar{C}_2 W=I_{f}$. This optimization problem can be solved using the Lagrange multiplier method \cite{lotte2010}, whose Lagrange function is
\begin{align}
F(W,\lambda)=W^{\top} \bar{C}_1 W-\lambda(W^{\top} \bar{C}_2 W-I_{f}).
\end{align}

Setting the derivative of $F(W,\lambda)$ with respect to $W$ to 0, we have
\begin{align}
\frac{\partial F(W,\lambda)}{\partial W}&= 2W^{\top}\bar{C}_1-2\lambda W^{\top}\bar{C}_2=0 \nonumber \\
&\Leftrightarrow \bar{C}_1 W=\lambda \bar{C}_2 W \nonumber \\
&\Leftrightarrow \bar{C}_2^{-1}\bar{C}_1 W=\lambda W, \nonumber
\end{align}
which becomes a standard eigenvalue decomposition problem.

The spatial filtering matrix $W$ consists of eigenvectors corresponding to the $\frac{f}{2}$ largest and the $\frac{f}{2}$ smallest eigenvalues of $\bar{C}_2^{-1}\bar{C}_1$.

\subsection{CNNs for MI Classification}

Five popular CNN models are considered in this paper: EEGNet~\cite{EEGNet}, DeepCNN~\cite{MNE}, ShallowCNN~\cite{MNE}, FBCNet~\cite{Mane2021}, and EEGConformer~\cite{Song2023}. Their architectures are detailed in Tables~\ref{tab:EEGNet}-\ref{tab:EEGConformer}, respectively.
\begin{itemize}
\item EEGNet~\cite{EEGNet}, which consists of three convolutional blocks and a classifier block. The first convolutional block performs temporal filtering for capturing frequency information. The second spatial filter block uses depthwise convolution with size $(c, 1)$ to learn spatial filters. The third separable convolutional block is used to reduce the number of parameters and decouple the relationships within and across feature maps.
\item DeepCNN~\cite{MNE}, compared with EEGNet, it is deeper and hence has much more parameters. It mainly includes a temporal convolutional block, a spatial filter block, two standard convolutional blocks and a classifier block. The first temporal and spatial convolutional blocks are specially designed to handle EEG inputs and the other two are standard ones.
\item ShallowCNN~\cite{MNE}, which is a shallow version of DeepCNN, inspired by FBCSP. Its first two blocks are similar to the temporal and spatial convolutional blocks of DeepCNN, but with a larger kernel, a different activation function, and a different pooling approach.
\item FBCNet~\cite{Mane2021}, which is a simple yet effective CNN architecture. It begins by applying multiple fixed-parameter band-pass filters to decompose the EEG into various frequency bands as multi-view inputs. Spatial filter block is then used to extract spatially discriminative patterns from each frequency band. Finally, a classifier block is designed for classification.
\item EEGConformer~\cite{Song2023}, which is a compact convolutional transformer model. The convolution module also includes a temporal convolutional block and a spatial filter block for learning the low-level local features. The multiple self-attention modules are used to extract the global correlation within the local features.
\end{itemize}

\begin{table}[htpb] \footnotesize
\centering \setlength{\tabcolsep}{3mm} \renewcommand{\arraystretch}{0.5}
\caption{EEGNet \cite{EEGNet}.}
\begin{tabular}{c|c|c|c}
\toprule
\multirow{2}[2]{*}{Block} & \multirow{2}[2]{*}{Layer} & \multirow{2}[2]{*}{Filter size} & Number of \\
          &       &       & filters \\ \midrule
Temporal & Conv2D & $(1, \frac{f_s}{2})$ & 4\\
convolution & Batch normalization & - & - \\ \midrule
\multirow{5}{*}  & DepthwiseConv2D & $(c, 1)$ & 8\\
 Depthwise & Batch normalization & - & - \\
 spatial filter & ELU activation & - & - \\
 & Average pooling & $(1, 4)$ & - \\
 & Dropout & - & - \\ \midrule
\multirow{7}{*} & SeparableConv2D & $(1, 16)$ &8\\
 & Batch normalization & - & - \\
 Separable & PointwiseCon2D & $(1, 1)$ & 8 \\
 convolution & Batch normalization & - & - \\
 & ELU activation & - & - \\
 & Average pooling & $(1, 8)$ & - \\
 & Dropout & - & - \\ \midrule
 Classifier & Fully connection & - & - \\ \bottomrule
\end{tabular}
\label{tab:EEGNet}
\end{table}

\begin{table}[htpb] \footnotesize
\centering \setlength{\tabcolsep}{3mm} \renewcommand{\arraystretch}{0.5}
\caption{DeepCNN \cite{MNE}.} 
\begin{tabular}{c|c|c|c}
\toprule
\multirow{2}[2]{*}{Block} & \multirow{2}[2]{*}{Layer} & \multirow{2}[2]{*}{Filter size} & Number of \\
          &       &       & filters \\ \midrule
 Temporal  & \multirow{2}{*}{Conv2D} & \multirow{2}{*}{$(1, 5)$} & \multirow{2}{*}{25} \\
 convolution & & & \\ \midrule
\multirow{5}{*}{Spatial filter}  & Conv2D & $(c, 1)$ & 25 \\
 & Batch normalization & - & - \\
 & ELU activation & - & - \\
 & Max pooling & $(1, 2)$ & - \\
 & Dropout & - & - \\ \midrule
 \multirow{5}{*} & Conv2D & $(1, 5)$ & 50 \\
 Standard & Batch normalization & - & - \\
 convolution & ELU activation & - & - \\
 & Max pooling & $(1, 2)$ & - \\
 & Dropout & - & - \\ \midrule
 \multirow{5}{*} & Conv2D & $(1, 5)$ & 100 \\
 Standard & Batch normalization & - & - \\
 convolution & ELU activation & - & - \\
 & Max pooling & $(1, 2)$ & - \\
 & Dropout & - & - \\ \midrule
 Classifier & Fully connection & - & - \\ \bottomrule
\end{tabular}
\label{tab:DeepCNN}
\end{table}

\begin{table}[htpb] \footnotesize
\centering \setlength{\tabcolsep}{3mm} \renewcommand{\arraystretch}{0.4}
\caption{ShallowCNN \cite{MNE}.}
\begin{tabular}{c|c|c|c}
\toprule
\multirow{2}[2]{*}{Block} & \multirow{2}[2]{*}{Layer} & \multirow{2}[2]{*}{Filter size} & Number of \\
          &       &       & filters \\ \midrule
 Temporal  & \multirow{2}{*}{Conv2D} & \multirow{2}{*}{$(1, 13)$} & \multirow{2}{*}{40} \\
 convolution & & & \\ \midrule
\multirow{6}{*}{Spatial filter}  & Conv2D & $(c, 1)$ & 40 \\
 & Batch Normalization & - & - \\
 & Squaring Activation & - & - \\
 & Average Pooling & $(1, 35)$ & - \\
 & Logarithmic Activation & - & - \\
 & Dropout & - & - \\ \midrule
 Classifier & Fully Connection & - & - \\ \bottomrule
\end{tabular}
\label{tab:ShallowCNN}
\end{table}

\begin{table}[htbp] \footnotesize
  \centering \setlength{\tabcolsep}{3mm} \renewcommand{\arraystretch}{0.5}
  \caption{FBCNet \cite{Mane2021}.}
    \begin{tabular}{c|c|c|c}
    \toprule
    \multirow{2}[2]{*}{Block} & \multirow{2}[2]{*}{Layer} & \multirow{2}[2]{*}{Filter size} & Number of \\
      &   &   & filters \\
    \midrule
    Band-pass & \multirow{2}[2]{*}{-} & \multirow{2}[2]{*}{-} & \multirow{2}[2]{*}{6} \\
    filter &   &   &  \\
    \midrule
      & DepthwiseConv2D & $(c,1)$ & 48 \\
    Spatial & Batch normalization & - & - \\
    filter & Swish activation & - & - \\
      & Variance layer & - & - \\
    \midrule
    Classifier & Fully connection & - & - \\
    \bottomrule
    \end{tabular}%
  \label{tab:FBCNet}%
\end{table}%

\begin{table}[htbp]
  \footnotesize
  \centering \setlength{\tabcolsep}{3mm} \renewcommand{\arraystretch}{0.5}
  \caption{EEGConformer \cite{Song2023}.}
    \begin{tabular}{c|c|c|c}
    \toprule
    \multirow{2}[2]{*}{Block} & \multirow{2}[2]{*}{Layer} & \multirow{2}[2]{*}{Filter size} & Number of \\
      &   &   & filters \\
    \midrule
    Temporal & \multirow{2}[2]{*}{Conv2D} & \multirow{2}[2]{*}{$(1,25)$} & \multirow{2}[2]{*}{40} \\
    convolution &   &   &  \\
    \midrule
      & Conv2D & $(c,1)$ & 40 \\
      & Batch normalization & - & - \\
      & ELU activation & - & - \\
    Spatial filter & Average pooling & $(1,75)$ & - \\
      & Dropout & - & - \\
      & PointwiseConv2D & $(1,1)$ & 40 \\
      & Rearrange & - & - \\
    \midrule
      & Layer normalization & - & - \\
      & MHA & - & - \\
      & Dropout & - & - \\
    6$\times$ & Residual add & - & - \\
    Self-attention & Layer normalization & - & - \\
      & FFN & - & - \\
      & Dropout & - & - \\
      & Residual add & - & - \\
    \bottomrule
    \end{tabular}%
  \label{tab:EEGConformer}%
\end{table}%

\subsection{CSP-Net-1}

Our proposed CSP-Net-1 simply performs CSP before a CNN.

\begin{figure*}[htpb]\centering
\includegraphics[width=0.9\linewidth,clip]{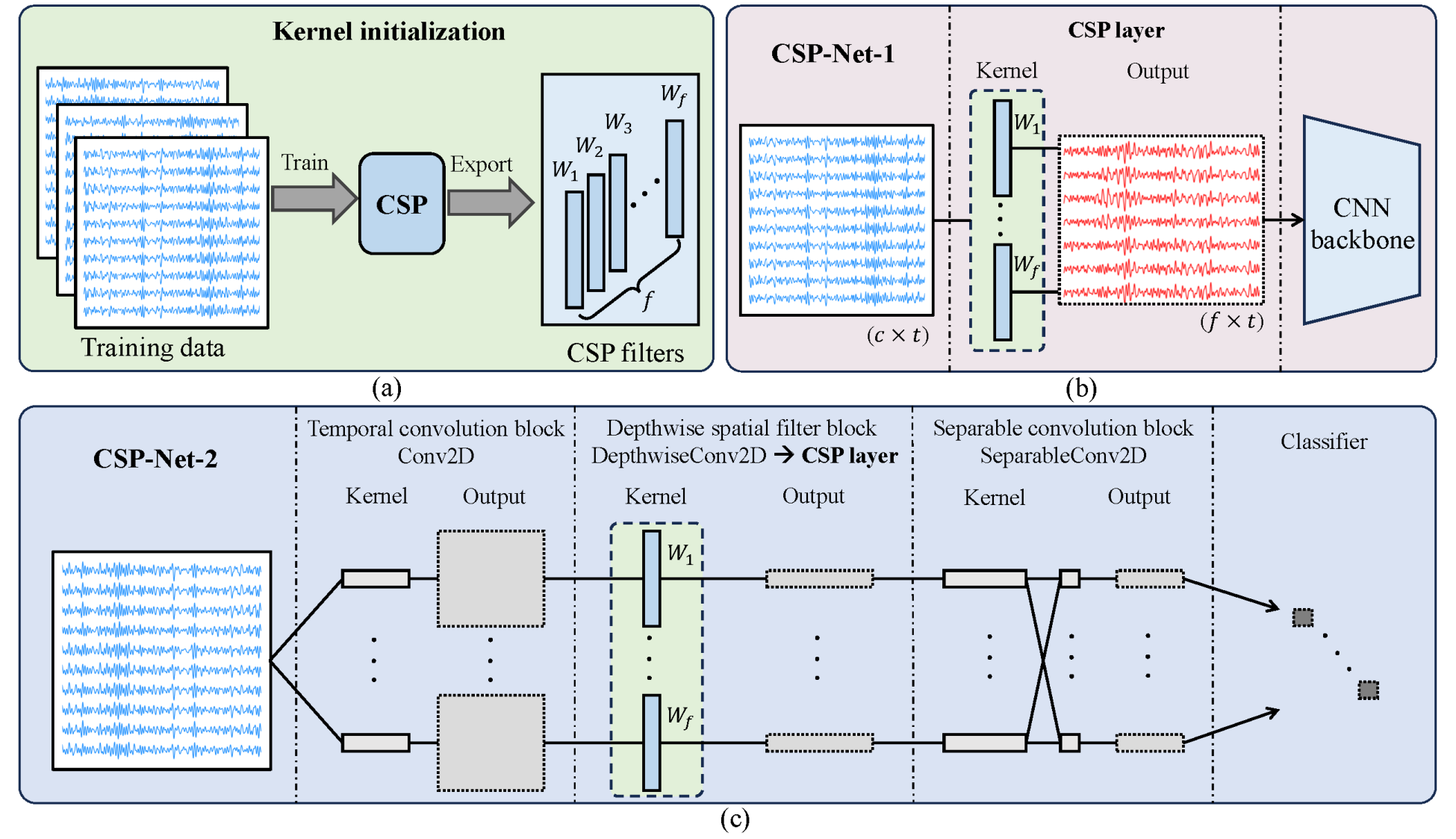}
\caption{Our proposed CSP-Nets. (a) Traditional CSP filters are used to initialize the CSP layer in CSP-Nets. (b) CSP-Net-1, which directly adds a CSP layer before a CNN backbone. (c) CSP-Net-2, illustrated using EEGNet \cite{EEGNet} (Table 1 in Supplementary Materials); the DepthwiseConv2D layer in its depthwise spatial filter block is replaced by a CSP layer.} \label{fig:framework}
\end{figure*}

As illustrated in Fig.~\ref{fig:framework}(a), all training EEG samples are used in CSP, resulting in $f$ spatial filters $W_i\in \mathbb{R}^{c\times 1}$, $i=1,\ldots,f$. Then, as shown in Fig.~\ref{fig:framework}(b), CSP-Net-1 uses these filters to spatially filter the raw EEG signals, before passing them to a CNN backbone.

There could be two different training approaches: 1) Fix the CSP layer and train the CNN backbone only (CSP-Net-1-fix); and, 2) update the CSP layer and the CNN backbone simultaneously (CSP-Net-1-upd). Their effectiveness will be discussed in Section~\ref{sect:results}.

CSP-Net-1 applies CSP filtering as a pre-processing step, enabling the model to work with more discriminative input signals. This explicit inclusion of the CSP filter provides a more structured way to embed expert knowledge into the network, thereby improving the model's capacity to capture task-relevant spatial features.

Algorithm~\ref{alg1} gives the pseudo-code of CSP-Net-1.
\begin{algorithm}
\DontPrintSemicolon
\KwIn{Training data $X$; a CNN model.\\}
\BlankLine
Perform CSP on $X$ to obtain the filter matrix $W$; \\
Initialize CSP-Net-1, which consists of a CSP layer with weights $W$ and a randomly initialized CNN;\\
Train CSP-Net-1 on $X$;\\
\BlankLine
\KwRet CSP-Net-1.
\caption{CSP-Net-1 for MI classification.}\label{alg1}
\end{algorithm}
\begin{algorithm}
\DontPrintSemicolon
\KwIn{Training data $X$; a CNN model.\\}
\BlankLine
Perform CSP on $X$ to obtain the filter matrix $W$;\\
Randomly initialize the CNN model;\\
Initialize CSP-Net-2, by replacing the convolutional layer in the spatial filter block of the CNN model by a CSP layer with weights $W$;\\
Train CSP-Net-2 on $X$;\\
\BlankLine
\KwRet CSP-Net-2.
\caption{CSP-Net-2 for MI classification.}\label{alg2}
\end{algorithm}
\subsection{CSP-Net-2}

Many CNN models have been proposed for MI classification, which typically consist of multiple convolution-pooling layers for feature extraction and some fully connected layers for classification. Although they differ in architecture, they usually include a spatial filter layer with spatial convolutional kernels specifically designed for EEG signals.

CSP-Net-2 replaces their spatial filter layer with a CSP layer. Fig.~\ref{fig:framework}(c) uses EEGNet as the CNN backbone to illustrate the architecture of CSP-Net-2. For clarity, we primarily depict the connection of the convolutional kernel between inputs and outputs. The depthwise spatial filter block aims to learn spatial patterns in EEG data. CSP-Net-2 replaces the convolutional kernels in this block with the CSP filters, and keeps other parts unchanged.

More specifically, CSP-Net-2 uses the CSP layer to replace the DepthwiseConv2D layer in the spatial filter block of EEGNet (8 kernels), the Conv2D layer in the spatial filter block of DeepCNN (25 kernels), the Conv2D layer in the spatial filter block of ShallowCNN (40 kernels), the DepthwiseConv2D layer in the spatial filter block of FBCNet (48 kernels), and the Conv2D layer in the spatial filter block of EEGConformer (40 kernels).

Similar to CSP-Net-1, the CSP filter layer in CSP-Net-2 can either be fixed (CSP-Net-2-fix) or updated (CSP-Net-2-upd). Furthermore, this replacement is significant as it allows CSP-Net-2 to explicitly incorporate prior knowledge about spatial filtering, enhancing the model's ability to capture discriminative features from the EEG signals. The flexibility of using either fixed or updated CSP filters also provides a balance between stability and adaptability during training, which we discussed in detail in Section~\ref{sect:results}.

Algorithm~\ref{alg2} gives the pseudo-code of CSP-Net-2.

\section{Experiments and Results} \label{sect:experiment}

This section presents the experimental results to validate the effectiveness of our proposed CSP-Nets.

\subsection{Datasets}

Four public MI datasets from BNCI-Horizon \footnote{http://www.bnci-horizon-2020.eu/database/data-sets}, summarized in Table~\ref{tab:dataset}, were used in our experiments:
\begin{enumerate}
  \item MI4C and MI2C: They were from the 001-2014 dataset. The EEG signals were sampled at 250Hz. MI2C includes only left-hand and right-hand trials. MI4C includes all classes.
  \item MI14S: This was from the 002-2014 dataset. The EEG signals were sampled at 512Hz.
  \item MI9S: This was the 001-2015 dataset. The EEG signals were recorded at 512Hz. The last three subjects were discarded due to their poor performance \cite{Faller2012,xia2022privacy}.
\end{enumerate}
They were downloaded and pre-processed using the MOABB framework \cite{jayaram2018moabb}. All datasets were pre-processed with an 8-32Hz bandpass filter.

\begin{table}[htpb]
  \centering  \small \setlength{\tabcolsep}{1mm}
  \caption{Summary of the four MI datasets.}
    \begin{tabular}{c|c|c|c|c}
    \toprule
    Datasets & \# Subjects & \# Channels & \# Trials per subject & \# Classes \\
    \midrule
    MI4C  & 9     & 22    & 288   & 4 \\
    MI2C  & 9     & 22    & 144   & 2 \\
    MI14S & 14    & 15    & 100   & 2 \\
    MI9S  & 9     & 13    & 200   & 2 \\
    \bottomrule
    \end{tabular}
  \label{tab:dataset}
\end{table}

\subsection{Implementation Details}

We evaluated the performance of CSP-Nets in both within-subject and cross-subject classifications:
\begin{enumerate}
  \item Within-subject classification: For each individual subject, 80\% trials were used for training, and the remaining 20\% for testing.
  \item Cross-subject classification: Leave-one-subject-out cross-validation was performed, i.e., one subject was used as the test set and all remaining ones as the training set.
\end{enumerate}

All experiments were repeated 5 times, and the average accuracies are reported.

We used Adam optimizer with batch size 128 and initial learning rate 0.01, and cross-entropy loss with weight decay 0.0005. The maximum number of training epochs was 200. The CSP layer used by default $f=8$ spatial filters (Section~\ref{sect:filter} presents sensitivity analysis). For CSP-Net-2, the number of convolutional kernels in the original spatial filter layer of the CNN models may be larger than 8. We expanded the CSP filters to address this mismatch: when the number of convolutional kernels exceeds the number of CSP filters, we replicate the CSP filters to match the number of required kernels. Specifically, we duplicated the 8 CSP filters 5 times to match the 40 kernels in the spatial filter block of ShallowCNN and EEGConformer ($8\times 5=40$), duplicated the 8 CSP filters 6 times to match the 48 kernels in the spatial filter block of FBCNet ($8\times 6=48$), and duplicated the 8 CSP filters 3 times and randomly selected one more to match the 25 kernels in the spatial filter block of DeepCNN ($8\times 3+1=25$).

\subsection{Experimental Results}\label{sect:results}

Table~\ref{tab:MI4C} shows the classification accuracies for the individual subjects on MI4C, where CSP-LR used logistic regression as the classifier. Tables~\ref{tab:MI2C_ave}-\ref{tab:MI9S_ave} show the average classification results across all subjects on the other three datasets, due to page limit. We performed paired $t$-tests on the results, calculated $p$-values between the standard backbone models and the CSP-Nets, and adjusted them using Benjamini Hochberg False Discovery Rate correction. Observe that:
\begin{enumerate}
  \item Both CSP-Nets were generally highly effective on all datasets and backbones. Embedding CSP knowledge in CNN backbones resulted in significant performance improvements. For example, in within-subject classification on MI4C, CSP-Net-2-fix increased the average accuracy on all subjects from 63.50\% to 71.91\% after integrating CSP information into EEGNet as CSP-Net-2-fix. The average accuracies across all five backbones also exhibited significant improvements, from an initial accuracy of 61.32\% to as high as 67.33\%.
  \item CSP-Nets with fixed CSP layer parameters generally performed better. Particularly, CSP-Net-2-fix achieved substantial improvements over EEGNet, DeepCNN, and EEGConformer. This validated that the incorporation of CSP prior knowledge can enhance the generalization of CNN models. High number and proportion of parameters of spatial convolutional kernels in ShallowCNN and FBCNet may overshadow the benefits offered by CSP filters.
  \item CSP-Nets had larger performance improvements in within-subject classification than cross-subject classification. This might be because: 1) within-subject classification had much fewer training samples than cross-subject classification, and hence prior knowledge in CSP is more helpful to the generalization performance; and, 2) cross-subject classification is intrinsically more challenging, as there are large individual differences among different subjects. The impact of training data quantity on CSP-Nets is discussed in Section~\ref{sect:smallsample}.
  \item CSP-Nets achieved better performance on most subjects. However, for some subjects where CSP did not perform well, CSP-Nets also struggled, e.g., Subject 1, 2, 5 and 6 in cross-subject classification on MI4C.
\end{enumerate}

\begin{table*}[htbp] \footnotesize
  \centering
  \caption{Classification accuracies (\%) on MI4C. Average accuracies higher than Standard are marked in bold. Asterisks indicate statistically significant differences between standard backbone and CSP-Net under adjusted paired $t$-test, where * means $p < 0.05$, ** means $p < 0.01$, *** means $p < 0.001$.} \setlength{\tabcolsep}{2mm}\renewcommand{\arraystretch}{0.95}
    \begin{tabular}{c|c|c|ccccccccc|c}
    \toprule
    Scenario & Backbone & Approach & S1    & S2    & S3    & S4    & S5    & S6    & S7    & S8    & S9    & \multicolumn{1}{c}{Average acc$\pm$std} \\
    \midrule
    &- & \multicolumn{1}{c}{CSP-LR} & 71.85  & 61.27  & 78.64  & 48.52  & 35.32  & 40.14  & 69.07  & 75.00  & 70.37  & \multicolumn{1}{c}{61.13} \\ \cmidrule{2-13}
    & \multirow{5}[2]{*}{EEGNet} & Standard & 72.59  & 49.35  & 81.36  & 44.51  & 49.31  & 39.29  & 65.98  & 84.19  & 84.93  & \multicolumn{1}{c}{63.50} \\
    & & CSP-Net-1-upd & 83.40  & 57.31  & 88.26  & 50.36  & 58.36  & 43.90  & 75.52  & 84.01  & 87.59  & 69.86***$\pm$1.54 \\
    & & CSP-Net-1-fix & 81.43  & 58.73  & 90.70  & 53.48  & 53.73  & 45.74  & 82.42  & 83.95  & 86.88  & 70.79***$\pm$1.68 \\
    & & CSP-Net-2-upd & 77.83  & 56.85  & 86.91  & 47.95  & 46.24  & 41.35  & 74.20  & 80.74  & 84.11  & 66.24$\pm$2.89 \\
    & & CSP-Net-2-fix & 81.18  & 63.55  & 92.36  & 52.69  & 56.75  & 46.27  & 83.02  & 82.64  & 88.76  & 71.91***$\pm$0.74 \\ \cmidrule{2-13}
    & \multirow{5}[2]{*}{DeepCNN} & Standard & 56.86  & 45.93  & 66.55  & 39.72  & 25.48  & 30.14  & 57.34  & 68.66  & 72.93  & 51.51$\pm$1.13 \\
    & & CSP-Net-1-upd & 69.07  & 57.07  & 80.32  & 50.76  & 38.34  & 37.82  & 68.72  & 73.76  & 78.98  & 61.65***$\pm$1.70 \\
    & & CSP-Net-1-fix & 70.12  & 56.24  & 81.90  & 52.06  & 46.05  & 38.28  & 69.04  & 72.94  & 82.85  & 63.28***$\pm$1.26 \\
    & & CSP-Net-2-upd & 62.52  & 41.36  & 59.02  & 43.28  & 25.64  & 25.00  & 54.47  & 70.01  & 71.74  & 50.34$\pm$0.95 \\
    & & CSP-Net-2-fix & 73.07  & 55.34  & 82.59  & 51.55  & 45.90  & 38.18  & 74.78  & 76.61  & 81.52  & 64.39***$\pm$2.46 \\ \cmidrule{2-13}
    & \multirow{5}[2]{*}{ShallowCNN} & Standard & 69.35  & 57.51  & 74.62  & 53.04  & 50.74  & 46.14  & 76.48  & 76.33  & 81.95  & 65.13$\pm$1.38 \\
    & & CSP-Net-1-upd & 79.99  & 57.93  & 84.78  & 57.87  & 54.98  & 44.48  & 85.43  & 80.79  & 80.92  & 69.69***$\pm$1.58 \\
    & & CSP-Net-1-fix & 80.15  & 59.93  & 86.51  & 58.04  & 52.83  & 44.48  & 85.14  & 83.04  & 83.64  & 70.42***$\pm$1.41 \\
    & & CSP-Net-2-upd & 70.21  & 57.53  & 79.94  & 50.79  & 37.88  & 40.42  & 76.08  & 83.20  & 80.67  & 64.08$\pm$1.62 \\
    Within- & & CSP-Net-2-fix & 69.04  & 63.02  & 86.55  & 45.62  & 31.48  & 38.89  & 77.06  & 79.06  & 82.27  & 63.66$\pm$2.01 \\ \cmidrule{2-13}
    Subject & \multirow{5}[2]{*}{FBCNet} & Standard & 69.27  & 53.87  & 83.77  & 50.17  & 53.03  & 43.78  & 69.61  & 80.22  & 83.95  & 65.30$\pm$1.76 \\
    & & CSP-Net-1-upd & 77.39  & 53.70  & 86.65  & 52.61  & 60.79  & 45.62  & 78.23  & 86.81  & 84.71  & 69.61***$\pm$1.23 \\
    & & CSP-Net-1-fix & 76.41  & 56.87  & 87.44  & 49.11  & 53.56  & 42.79  & 77.72  & 86.07  & 86.01  & 68.44**$\pm$2.04 \\
    & & CSP-Net-2-upd & 71.19  & 54.88  & 87.29  & 53.58  & 54.56  & 43.96  & 72.34  & 83.05  & 86.46  & 67.48*$\pm$1.85 \\
    & & CSP-Net-2-fix & 76.74  & 57.42  & 86.45  & 46.79  & 53.81  & 43.64  & 71.47  & 82.65  & 85.98  & 67.22$\pm$1.02 \\ \cmidrule{2-13}
    & \multirow{5}[2]{*}{EEGConformer} & Standard & 75.00  & 46.97  & 79.81  & 50.23  & 37.46  & 46.05  & 72.25  & 77.19  & 65.42  & 61.15$\pm$1.67 \\
    & & CSP-Net-1-upd & 81.98  & 63.81  & 87.68  & 58.57  & 60.62  & 52.46  & 88.04  & 82.96  & 71.73  & 71.98***$\pm$1.38 \\
    & & CSP-Net-1-fix & 84.28  & 59.84  & 87.41  & 61.85  & 58.49  & 53.90  & 90.18  & 79.96  & 71.82  & 71.97***$\pm$1.14 \\
    & & CSP-Net-2-upd & 80.49  & 58.97  & 89.79  & 60.50  & 56.39  & 55.85  & 84.24  & 85.14  & 75.04  & 71.82***$\pm$1.04 \\
    & & CSP-Net-2-fix & 82.55  & 60.73  & 86.94  & 45.70  & 58.18  & 46.37  & 78.67  & 83.13  & 82.95  & 69.47***$\pm$1.04 \\ \cmidrule{2-13}
    & \multirow{5}[1]{*}{Average} & Standard & 68.61  & 50.73  & 77.22  & 47.53  & 43.20  & 41.08  & 68.33  & 77.32  & 77.84  & 61.32$\pm$1.57 \\
    & & CSP-Net-1-upd & \textbf{78.37} & \textbf{57.96} & \textbf{85.54} & \textbf{54.03} & \textbf{54.62} & \textbf{44.86} & \textbf{79.19} & \textbf{81.67} & \textbf{80.79} & \textbf{68.56}***$\pm$1.49 \\
    & & CSP-Net-1-fix & \textbf{78.48} & \textbf{58.32} & \textbf{86.79} & \textbf{54.91} & \textbf{52.93} & \textbf{45.04} & \textbf{80.90} & \textbf{81.19} & \textbf{82.24} & \textbf{68.98}***$\pm$1.51 \\
    & & CSP-Net-2-upd & \textbf{72.45} & \textbf{53.92} & \textbf{80.59} & \textbf{51.22} & \textbf{44.14} & \textbf{41.32} & \textbf{72.27} & \textbf{80.43} & \textbf{79.60} & \textbf{63.99}***$\pm$1.67 \\
    & & CSP-Net-2-fix & \textbf{76.52} & \textbf{60.01} & \textbf{86.98} & \textbf{48.47} & \textbf{49.22} & \textbf{42.67} & \textbf{77.00} & \textbf{80.82} & \textbf{84.30} & \textbf{67.33}***$\pm$1.45 \\
    \midrule
    &- & \multicolumn{1}{c}{CSP-LR} & 61.46  & 22.92  & 72.22  & 43.06  & 32.29  & 39.58  & 62.50  & 76.04  & 62.85  & \multicolumn{1}{c}{52.55} \\ \cmidrule{2-13}
    & \multirow{5}[2]{*}{EEGNet} & Standard & 68.96  & 30.35  & 71.88  & 38.06  & 37.01  & 36.74  & 42.01  & 58.19  & 61.74  & 49.44$\pm$1.75 \\
    & & CSP-Net-1-upd & 65.62  & 32.15  & 72.92  & 41.39  & 34.93  & 38.68  & 55.28  & 65.69  & 61.74  & 52.04*$\pm$1.48 \\
    & & CSP-Net-1-fix & 62.50  & 32.43  & 76.32  & 40.97  & 34.58  & 37.43  & 52.01  & 67.64  & 65.62  & 52.17*$\pm$1.13 \\
    & & CSP-Net-2-upd & 66.94  & 32.64  & 74.17  & 42.01  & 35.07  & 40.97  & 42.78  & 67.08  & 63.82  & 51.72*$\pm$1.59 \\
    & & CSP-Net-2-fix & 66.46  & 31.53  & 74.44  & 39.65  & 31.04  & 37.29  & 53.40  & 63.06  & 68.61  & 51.72$\pm$1.38 \\ \cmidrule{2-13}
    & \multirow{5}[2]{*}{DeepCNN} & Standard & 65.35  & 34.03  & 52.99  & 37.92  & 38.19  & 43.40  & 41.81  & 59.24  & 53.47  & 47.38$\pm$2.10 \\
    & & CSP-Net-1-upd & 61.32  & 33.40  & 67.57  & 41.81  & 35.56  & 40.90  & 43.19  & 63.19  & 57.01  & 49.33$\pm$0.70 \\
    & & CSP-Net-1-fix & 60.07  & 34.79  & 62.22  & 42.50  & 35.21  & 40.97  & 41.67  & 67.01  & 60.14  & 49.40$\pm$0.53 \\
    & & CSP-Net-2-upd & 64.93  & 31.18  & 65.00  & 38.96  & 39.58  & 41.67  & 48.61  & 56.94  & 57.15  & 49.34$\pm$1.11 \\
    & & CSP-Net-2-fix & 62.01  & 29.93  & 69.44  & 37.99  & 36.32  & 39.44  & 52.92  & 63.33  & 61.67  & 50.34*$\pm$1.81 \\ \cmidrule{2-13}
    &  \multirow{5}[2]{*}{ShallowCNN} & Standard & 68.12  & 33.40  & 71.04  & 41.46  & 36.18  & 45.49  & 43.82  & 69.51  & 64.17  & 52.58$\pm$0.65 \\
    & & CSP-Net-1-upd & 66.25  & 30.63  & 70.69  & 41.53  & 32.15  & 40.49  & 50.00  & 64.65  & 59.03  & 50.60$\pm$1.23 \\
    & & CSP-Net-1-fix & 68.12  & 31.04  & 72.01  & 42.71  & 32.78  & 39.24  & 53.96  & 69.38  & 61.60  & 52.31$\pm$1.28 \\
    & & CSP-Net-2-upd & 66.39  & 30.63  & 72.57  & 43.12  & 32.29  & 43.06  & 47.29  & 67.57  & 63.61  & 51.84$\pm$1.25 \\
    Cross- & & CSP-Net-2-fix & 62.15  & 28.47  & 72.08  & 39.58  & 32.36  & 41.67  & 54.93  & 67.64  & 67.64  & 51.84$\pm$1.04 \\ \cmidrule{2-13}
    Subject & \multirow{5}[2]{*}{FBCNet} & Standard & 62.92  & 31.39  & 62.22  & 41.67  & 31.25  & 36.25  & 40.76  & 55.21  & 57.01  & 46.52$\pm$1.49 \\
    & & CSP-Net-1-upd & 66.18  & 30.90  & 62.50  & 42.43  & 31.53  & 38.33  & 43.26 & 57.43 & 60.49  & 48.12$\pm$1.31 \\
    & & CSP-Net-1-fix & 67.22  & 31.39  & 67.43  & 39.51  & 31.11  & 38.06  & 45.90  & 60.76  & 60.28  & 49.07**$\pm$1.11 \\
    & & CSP-Net-2-upd & 64.65  & 30.42  & 61.94  & 41.39  & 30.83  & 37.92  & 40.69  & 56.88  & 57.64  & 46.93$\pm$1.27 \\
    & & CSP-Net-2-fix & 57.50  & 32.50  & 64.38  & 40.35  & 36.67  & 36.88  & 49.79  & 64.31  & 65.00  & 49.71**$\pm$1.24 \\ \cmidrule{2-13}
    & \multirow{5}[2]{*}{EEGConformer} & Standard & 52.64  & 27.64  & 49.51  & 33.68  & 32.29  & 39.86  & 30.14  & 54.51  & 41.60  & 40.21$\pm$1.67 \\
    & & CSP-Net-1-upd & 61.94  & 35.07  & 63.06  & 34.38  & 32.36  & 37.85  & 26.11  & 61.39  & 55.76  & 45.32***$\pm$0.90 \\
    & & CSP-Net-1-fix & 58.54  & 35.83  & 63.33  & 35.35  & 32.64  & 38.19  & 27.78  & 64.03  & 56.18  & 45.76***$\pm$1.09 \\
    & & CSP-Net-2-upd & 55.76  & 33.06  & 71.67  & 38.19  & 33.19  & 35.14  & 44.03  & 61.39  & 63.75  & 48.46***$\pm$0.62 \\
    & & CSP-Net-2-fix & 56.81  & 33.26  & 72.92  & 39.65  & 31.11  & 37.57  & 44.44  & 67.22  & 65.83  & 49.87***$\pm$0.94 \\ \cmidrule{2-13}
    & \multirow{5}[1]{*}{Average} & Standard & 63.60  & 31.36  & 61.53  & 38.56  & 34.98  & 40.35  & 39.71  & 59.33  & 55.60  & 47.22$\pm$1.53 \\
    & & CSP-Net-1-upd & \textbf{64.26} & \textbf{32.43} & \textbf{67.35} & \textbf{40.31} & 33.31  & 39.25  & \textbf{43.57} & \textbf{62.47} & \textbf{58.81} & \textbf{49.08}***$\pm$1.12 \\
    & & CSP-Net-1-fix & 63.29  & \textbf{33.10} & \textbf{68.26} & \textbf{40.21} & 33.26  & 38.78  & \textbf{44.26} & \textbf{65.76} & \textbf{60.76} & \textbf{49.74}***$\pm$1.03 \\
    & & CSP-Net-2-upd & 63.73  & 31.59  & \textbf{69.07} & \textbf{40.73} & 34.19  & 39.75  & \textbf{44.68} & \textbf{61.97} & \textbf{61.19} & \textbf{49.66}***$\pm$1.17 \\
    & & CSP-Net-2-fix & 60.99  & 31.14  & \textbf{70.65} & \textbf{39.44} & 33.50  & 38.57  & \textbf{51.10} & \textbf{65.11} & \textbf{65.75} & \textbf{50.69}***$\pm$1.28 \\ \bottomrule
    \end{tabular}%
  \label{tab:MI4C}%
\end{table*}%

\begin{table*}[htbp] \scriptsize
  \centering \centering \setlength{\tabcolsep}{3.5mm} \renewcommand{\arraystretch}{1.0}
  \caption{Average classification accuracies (\%) on MI2C. Those higher than Standard are marked in bold. Asterisks indicate statistically significant differences between standard backbone and CSP-Net under adjusted paired $t$-test, where * means $p < 0.05$, ** means $p < 0.01$, *** means $p < 0.001$.}
\begin{tabular}{c|c|cccccc}
\toprule
\multirow{2}{*}{Scenario} & \multirow{2}{*}{Approach} & \multicolumn{6}{c}{Backbone} \\ \cmidrule{3-8}
&    & EEGNet & DeepCNN & ShallowCNN & FBCNet & EEGConformer & Average acc$\pm$std \\ \midrule
& CSP-LR   & \multicolumn{1}{c}{-} & \multicolumn{1}{c}{-} & \multicolumn{1}{c}{-} & \multicolumn{1}{c}{-} & \multicolumn{1}{c}{-} & 75.72  \\ \cmidrule{2-8}
& Standard & 76.38$\pm$2.21 & 61.46$\pm$3.28 & 76.29$\pm$2.92 & 78.11$\pm$2.64 & 75.31$\pm$1.45 & 73.51$\pm$2.50 \\
Within- & CSP-Net-1-upd & \textbf{80.02}$\pm$2.80 & \textbf{70.86}***$\pm$3.58 & \textbf{82.50}**$\pm$2.60 & \textbf{80.70}*$\pm$1.85 & \textbf{81.06}***$\pm$1.46 & \textbf{79.02}***$\pm$2.45 \\
subject & CSP-Net-1-fix & \textbf{81.69}*$\pm$0.49 & \textbf{70.37}***$\pm$3.41 & \textbf{83.71}***$\pm$1.45 & \multicolumn{1}{c}{\textbf{82.39}**$\pm$2.70} & \textbf{82.05}***$\pm$0.88 & \textbf{80.04}***$\pm$1.78 \\
& CSP-Net-2-upd & 75.94$\pm$2.24 & \textbf{61.59}$\pm$1.93 & \textbf{77.33}$\pm$1.68 & \textbf{79.65}$\pm$2.48 & \textbf{81.53}***$\pm$2.80 & \textbf{75.21}**$\pm$2.23 \\
& CSP-Net-2-fix & \textbf{79.66}*$\pm$2.38 & \textbf{75.86}***$\pm$1.11 & 75.34$\pm$1.81 & \textbf{79.54}$\pm$0.58 & \textbf{81.18}**$\pm$1.61 & \textbf{78.32}***$\pm$1.50 \\\midrule
& CSP-LR   & \multicolumn{1}{c}{-} & \multicolumn{1}{c}{-} & \multicolumn{1}{c}{-} & \multicolumn{1}{c}{-} & \multicolumn{1}{c}{-} & 72.92  \\
\cmidrule{2-8}
& Standard & \textbf{71.50}$\pm$0.87 & 73.15$\pm$1.37 & 74.32$\pm$1.50 & 69.34$\pm$1.50 & 65.25$\pm$1.03 & 70.71$\pm$1.16 \\
Cross- & CSP-Net-1-upd & \textbf{73.53}*$\pm$1.39 & \textbf{74.86}*$\pm$0.73 & \textbf{74.65}$\pm$0.63 & \textbf{71.00}$\pm$1.60 & \textbf{70.94}***$\pm$0.67 & \textbf{73.00}***$\pm$1.00 \\
subject & CSP-Net-1-fix & \textbf{74.51}**$\pm$1.28 & \textbf{74.75}$\pm$1.36 & \textbf{75.51}$\pm$0.65 & \textbf{73.09}***$\pm$1.35 & \textbf{71.20}***$\pm$0.57 & \textbf{73.81}***$\pm$1.04 \\
& CSP-Net-2-upd & \textbf{72.31}$\pm$1.23 & 72.93$\pm$0.48 & 70.40$\pm$0.87 & \textbf{69.41}$\pm$1.71 & \textbf{70.76}***$\pm$1.37 & \textbf{71.16}$\pm$1.13 \\
& CSP-Net-2-fix & \textbf{75.25}***$\pm$1.28 & \textbf{73.38}$\pm$0.91 & \textbf{76.11}$\pm$0.80 & \textbf{73.61}***$\pm$0.27 & \textbf{72.95}***$\pm$0.55 & \textbf{74.26}***$\pm$0.76 \\\bottomrule
\end{tabular}%
\label{tab:MI2C_ave}%
\end{table*}%

\begin{table*}[htbp]\scriptsize
  \centering \setlength{\tabcolsep}{3.5mm} \renewcommand{\arraystretch}{1.0}
\caption{Average classification accuracies (\%) on MI14S. Those higher than Standard are marked in bold. Asterisks indicate statistically significant differences between standard backbone and CSP-Net under adjusted paired $t$-test, where * means $p < 0.05$, ** means $p < 0.01$, *** means $p < 0.001$.}
\begin{tabular}{c|c|cccccc}
\toprule
\multirow{2}[4]{*}{Scenario} & \multirow{2}[4]{*}{Approach} & \multicolumn{6}{c}{Backbone} \\
\cmidrule{3-8}
&       & \multicolumn{1}{c}{EEGNet} & \multicolumn{1}{c}{DeepCNN} & \multicolumn{1}{c}{ShallowCNN} & \multicolumn{1}{c}{FBCNet} & \multicolumn{1}{c}{EEGConformer} & Average acc$\pm$std \\
    \midrule
& CSP-LR   & \multicolumn{1}{c}{-} & \multicolumn{1}{c}{-} & \multicolumn{1}{c}{-} & \multicolumn{1}{c}{-} & \multicolumn{1}{c}{-} & 74.95  \\\cmidrule{2-8}
& Standard & 75.22$\pm$1.94 & 55.75$\pm$2.92 & 70.70$\pm$1.90 & 78.55$\pm$1.75 & 75.72$\pm$1.88 & 71.19$\pm$2.08 \\
Within- & CSP-Net-1-upd & \textbf{76.69}$\pm$3.23 & \textbf{61.03}**$\pm$3.12 & \textbf{73.12}$\pm$2.02 & \multicolumn{1}{c}{\textbf{81.25}*$\pm$1.50} & \textbf{81.07}**$\pm$2.21 & \textbf{74.6}***$\pm$2.42 \\
subject & CSP-Net-1-fix & \textbf{78.92}*$\pm$1.67 & \textbf{61.33}**$\pm$1.51 & \textbf{74.89}*$\pm$1.84 & \multicolumn{1}{c}{\textbf{81.17}$\pm$1.98} & \textbf{80.96}**$\pm$1.23 & \textbf{75.45}***$\pm$1.61 \\
& CSP-Net-2-upd & \textbf{76.61}$\pm$2.37 & \textbf{56.84}$\pm$1.19 & \textbf{73.94}**$\pm$1.35 & \textbf{79.43}$\pm$2.90 & \textbf{79.96}*$\pm$2.32 & \textbf{73.36}***$\pm$2.03 \\
& CSP-Net-2-fix & \textbf{80.01}**$\pm$1.47 & \textbf{66.81}***$\pm$2.26 & \textbf{77.14}***$\pm$2.17 & 77.51$\pm$2.06 & \textbf{78.64}$\pm$3.25 & \textbf{76.02}***$\pm$2.25 \\
    \midrule
& CSP-LR   & \multicolumn{1}{c}{-} & \multicolumn{1}{c}{-} & \multicolumn{1}{c}{-} & \multicolumn{1}{c}{-} & \multicolumn{1}{c}{-} & 74.21  \\\cmidrule{2-8}
& Standard & 73.37$\pm$1.09 & 68.51$\pm$1.26 & 70.80$\pm$0.48 & 72.91$\pm$0.91 & 63.10$\pm$2.41 & 69.74$\pm$1.23 \\
Cross- & CSP-Net-1-upd & 73.19$\pm$1.34 & \textbf{71.94}***$\pm$1.26 & \textbf{73.03}**$\pm$0.22 & \multicolumn{1}{c}{\textbf{73.08}$\pm$0.93} & \textbf{73.83}***$\pm$0.61 & \textbf{73.01}***$\pm$0.76 \\
subject & CSP-Net-1-fix & \textbf{76.60}***$\pm$0.62 & \textbf{71.56}***$\pm$0.55 & \textbf{73.87}***$\pm$0.30 & \multicolumn{1}{c}{\textbf{73.91}$\pm$0.55} & \textbf{73.71}***$\pm$0.47 & \textbf{73.93}***$\pm$0.49 \\
& CSP-Net-2-upd & 73.07$\pm$1.94 & \textbf{70.04}$\pm$1.28 & 69.67$\pm$1.43 & \textbf{73.11}$\pm$1.64 & \textbf{70.93}***$\pm$1.06 & \textbf{71.36}***$\pm$1.47 \\
& CSP-Net-2-fix & \textbf{75.44}*$\pm$1.15 & \textbf{71.06}***$\pm$0.31 & 69.83$\pm$0.97 & \textbf{72.64}$\pm$0.70 & \textbf{71.21}***$\pm$0.48 & \textbf{72.04}***$\pm$0.72 \\\bottomrule
\end{tabular}%
\label{tab:MI14S_ave}%
\end{table*}%

\begin{table*}[htbp]\scriptsize
  \centering \setlength{\tabcolsep}{3.5mm} \renewcommand{\arraystretch}{1.0}
  \caption{Average classification accuracies (\%) on MI9S. Those higher than Standard are marked in bold. Asterisks indicate statistically significant differences between standard backbone and CSP-Net under adjusted paired $t$-test, where * means $p < 0.05$, ** means $p < 0.01$, *** means $p < 0.001$.}
\begin{tabular}{c|c|cccccc}
\toprule
\multirow{2}[4]{*}{Scenario} & \multirow{2}[4]{*}{Approach} & \multicolumn{6}{c}{Backbone} \\\cmidrule{3-8}
&       & \multicolumn{1}{c}{EEGNet} & \multicolumn{1}{c}{DeepCNN} & \multicolumn{1}{c}{ShallowCNN} & \multicolumn{1}{c}{FBCNet} & \multicolumn{1}{c}{EEGConformer} & \multicolumn{1}{c}{Average acc$\pm$std} \\
    \midrule
& CSP-LR   & \multicolumn{1}{c}{-} & \multicolumn{1}{c}{-} & \multicolumn{1}{c}{-} & \multicolumn{1}{c}{-} & \multicolumn{1}{c}{-} & \multicolumn{1}{c}{67.84} \\
\cmidrule{2-8}
& Standard & 70.85$\pm$0.88 & 62.34$\pm$3.39 & 69.03$\pm$1.51 & 72.55$\pm$1.49 & 69.65$\pm$1.08 & \multicolumn{1}{c}{68.88$\pm$1.67} \\
Within- & CSP-Net-1-upd & \textbf{73.01}$\pm$2.31 & \textbf{62.92}$\pm$1.53 & \textbf{72.59}*$\pm$0.98 & \multicolumn{1}{c}{\textbf{73.70}$\pm$1.12} & \textbf{74.91}**$\pm$2.22 & \multicolumn{1}{c}{\textbf{71.43}***$\pm$1.63} \\
subject & CSP-Net-1-fix & \textbf{73.63}*$\pm$1.85 & \textbf{63.63}$\pm$1.78 & \textbf{73.17}***$\pm$1.43 & \multicolumn{1}{c}{\textbf{74.31}$\pm$2.17} & \textbf{73.22}*$\pm$0.79 & \multicolumn{1}{c}{\textbf{71.59}***$\pm$$\pm$1.61} \\
& CSP-Net-2-upd & \textbf{72.01}$\pm$3.06 & \textbf{62.78}$\pm$3.09 & \textbf{71.38}*$\pm$1.07 & 71.67$\pm$2.49 & \textbf{73.55}*$\pm$2.07 & \multicolumn{1}{c}{\textbf{70.28}**$\pm$2.36} \\
& CSP-Net-2-fix & \textbf{71.83}$\pm$1.98 & \textbf{65.57}$\pm$0.96 & 67.19$\pm$1.68 & 72.41$\pm$2.57 & \textbf{72.79}$\pm$2.76 & \multicolumn{1}{c}{\textbf{69.96}$\pm$1.99} \\\midrule
& CSP-LR   & \multicolumn{1}{c}{-} & \multicolumn{1}{c}{-} & \multicolumn{1}{c}{-} & \multicolumn{1}{c}{-} & \multicolumn{1}{c}{-} & \multicolumn{1}{c}{57.72} \\
\cmidrule{2-8}
& Standard & 62.40$\pm$1.26 & 54.99$\pm$1.15 & 62.26$\pm$0.79 & 63.84$\pm$1.35 & 58.92$\pm$1.74 & 60.48$\pm$1.26 \\
Cross- & CSP-Net-1-upd & \textbf{64.77}*$\pm$1.75 & \textbf{59.49}**$\pm$2.51 & 61.73$\pm$0.70 & \multicolumn{1}{c}{63.41}$\pm$0.69 & \textbf{62.73}**$\pm$0.92 & \textbf{62.43}***$\pm$1.31 \\
subject & CSP-Net-1-fix & \textbf{64.91}*$\pm$2.28 & \textbf{60.07}**$\pm$1.91 & 62.16$\pm$1.45 & \multicolumn{1}{c}{62.88}$\pm$0.95 & \textbf{63.11}**$\pm$0.59 & \textbf{62.63}**$\pm$1.44 \\
& CSP-Net-2-upd & 62.03$\pm$1.76 & \textbf{61.98}***$\pm$0.95 & \textbf{63.53}$\pm$1.78 & \textbf{63.54}$\pm$1.15 & \textbf{59.96}$\pm$0.91 & \textbf{62.21}**$\pm$1.31 \\
& CSP-Net-2-fix & \textbf{64.72}$\pm$0.59 & \textbf{59.87}**$\pm$1.70 & 62.14$\pm$0.58 & 61.14$\pm$1.49 & \textbf{59.96}$\pm$0.69 & \textbf{61.57}$\pm$1.01 \\
\bottomrule
\end{tabular}%
\label{tab:MI9S_ave}%
\end{table*}%

\subsection{Comparative Performance Analysis}

We further compared our approaches with nine other approaches, including the state-of-the-art traditional approaches and deep learning approaches: CSP~\cite{ramoser2000}, FBCSP~\cite{FBCSP}, MDRM~\cite{Barachant2012}, DeepCNN~\cite{MNE}, LMDA-Net~\cite{miao2023lmda}, ShallowCNN~\cite{MNE}, EEGConformer~\cite{Song2023}, EEGNet~\cite{EEGNet}, and FBCNet~\cite{Mane2021}. Table~\ref{tab:comparison} presents the classification accuracies of CSP-Net-1 and CSP-Net-2 compared to these baselines. In CSP-Net-1 and CSP-Net-2, the EEGNet was used as the backbone architecture, and the fixed CSP layer was applied. The same training and test data were used for all models.

Both CSP-Net-1 and CSP-Net-2 demonstrated superior performance compared to traditional approaches like CSP and FBCSP, as well as more recent models like FBCNet and EEGConformer. These results highlight their effectiveness for EEG signal classification tasks.

\begin{table*}[htbp]\scriptsize
  \centering
  \caption{Classification accuracies (\%) for each model, averaged over all subjects.} \setlength{\tabcolsep}{2mm}
    \begin{tabular}{c|c|ccccccccc|cc}
    \toprule
    \multirow{2}[2]{*}{Scenario} &\multirow{2}[2]{*}{Dataset} & CSP- & FBCSP- & \multirow{2}[2]{*}{MDRM} & Deep- & LMDA- & Shallow- & EEGCon- & \multirow{2}[2]{*}{EEGNet} & \multirow{2}[2]{*}{FBCNet} & \multirow{2}[2]{*}{CSP-Net-1} & \multirow{2}[2]{*}{CSP-Net-2} \\
    &   &LR   &LR   &   & CNN & Net & CNN & former &   &   &   &  \\
    \midrule
    & MI4C & 61.43  & 66.81  & 64.21  & 51.51  & 56.75  & 65.13  & 61.15  & 63.50  & 65.30  & 71.98  & 71,82 \\
    & MI2C & 75.72  & 76.72  & 76.32  & 61.46  & 71.19  & 76.29  & 75.31  & 76.38  & 78.11  & 81.06  & 81.53  \\
    Within-subject & MI14S & 74.95  & 75.39  & 78.58  & 55.75  & 72.04  & 70.70  & 75.72  & 75.22  & 78.55  & 81.07  & 79.96  \\
    & MI9S & 67.84  & 69.37  & 67.90  & 62.34  & 69.94  & 69.03  & 69.65  & 70.85  & 72.55  & 74.91  & 73.55  \\
    \cmidrule{2-13}
    & Average & 69.99  & 72.07  & 71.75  & 57.77  & 67.48  & 70.29  & 70.46  & 71.49  & 73.63  & 76.26  & 75.85  \\
    \midrule
    & MI4C & 52.55  & 49.72  & 51.08  & 47.38  & 48.28  & 52.58  & 40.21  & 49.44  & 46.52  & 52.17  & 51.72  \\
    & MI2C & 72.92  & 73.62  & 71.99  & 73.15  & 69.10  & 74.32  & 65.25  & 71.50  & 69.34  & 74.51  & 75.25  \\
    Cross-subject & MI14S & 74.21  & 76.36  & 71.71  & 68.51  & 67.87  & 70.80  & 63.10  & 73.37  & 72.91  & 76.60  & 75.44  \\
    & MI9S & 57.27  & 63.61  & 57.39  & 54.99  & 63.41  & 62.26  & 58.92  & 62.40  & 63.84  & 64.91  & 64.72  \\
\cmidrule{2-13}
    & Average & 64.24  & 65.83  & 63.04  & 61.01  & 62.17  & 64.99  & 56.87  & 64.18  & 63.15  & 67.05  & 66.78  \\
    \bottomrule
    \end{tabular}%
  \label{tab:comparison}%
\end{table*}%

\subsection{Small Sample Setting} \label{sect:smallsample}

\begin{figure*}[htbp]\centering
\subfigure[]{\label{fig:MI4C_within}  \includegraphics[width=\linewidth,clip]{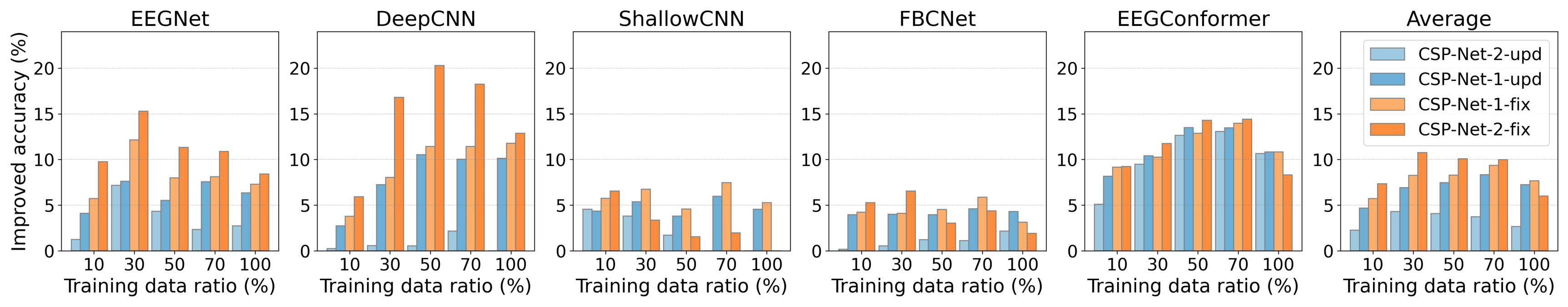}}
\subfigure[]{\label{fig:MI4C_cross}   \includegraphics[width=\linewidth,clip]{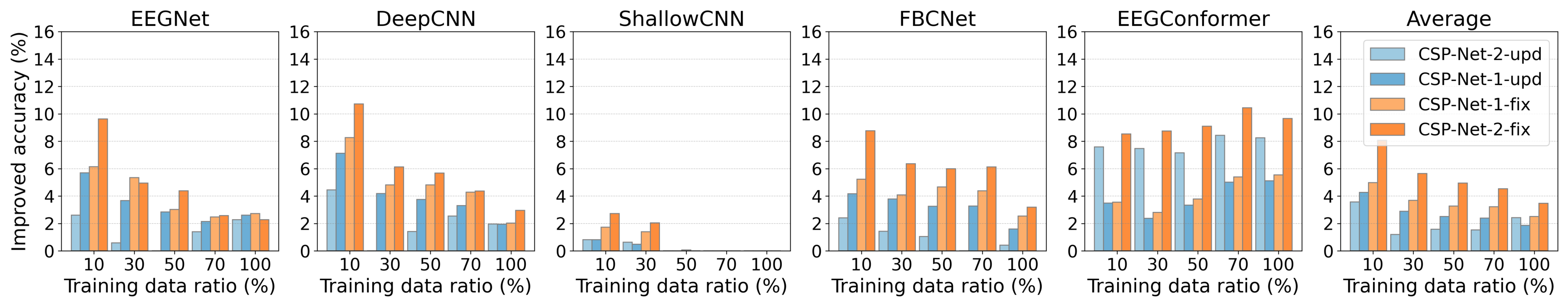}}
\caption{Accuracy improvements of CSP-Nets at different training data ratios on MI4C. (a) within-subject classification; and, (b) cross-subject classification.} \label{fig:data_ratio_MI4C}
\end{figure*}

\begin{figure*}[htbp]\centering
\subfigure[]{\label{fig:MI2C_within}  \includegraphics[width=\linewidth,clip]{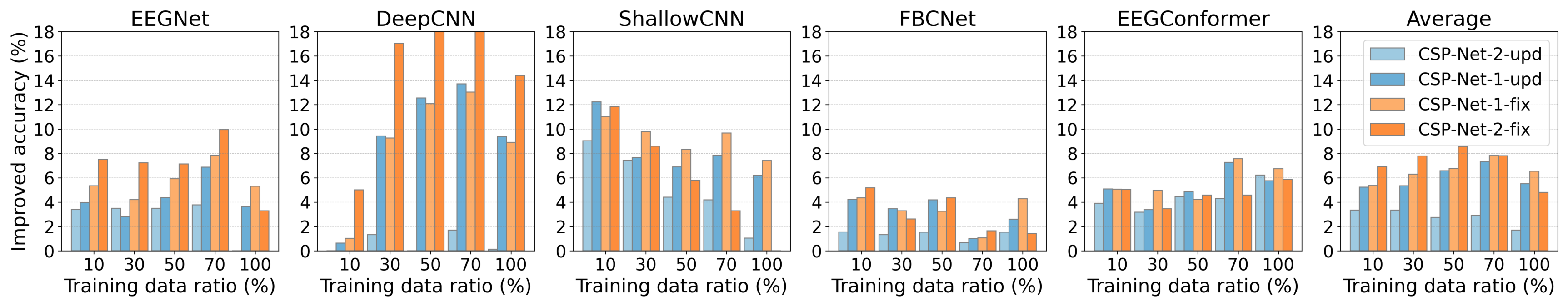}}
\subfigure[]{\label{fig:MI2C_cross}   \includegraphics[width=\linewidth,clip]{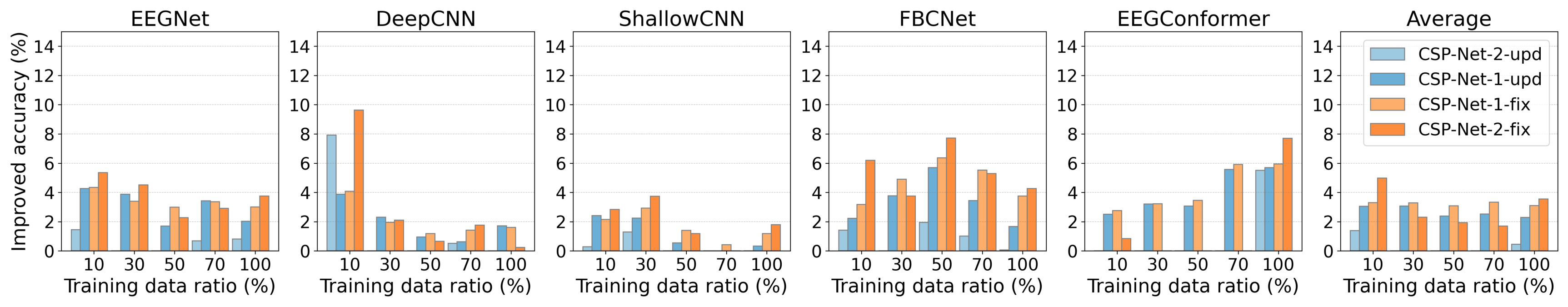}}
\caption{Accuracy improvements of CSP-Nets at different training data ratios on MI2C. (a) within-subject classification; and, (b) cross-subject classification.} \label{fig:data_ratio_MI2C}
\end{figure*}

\begin{figure*}[htbp]\centering
\subfigure[]{\label{fig:MI14S_within} \includegraphics[width=\linewidth,clip]{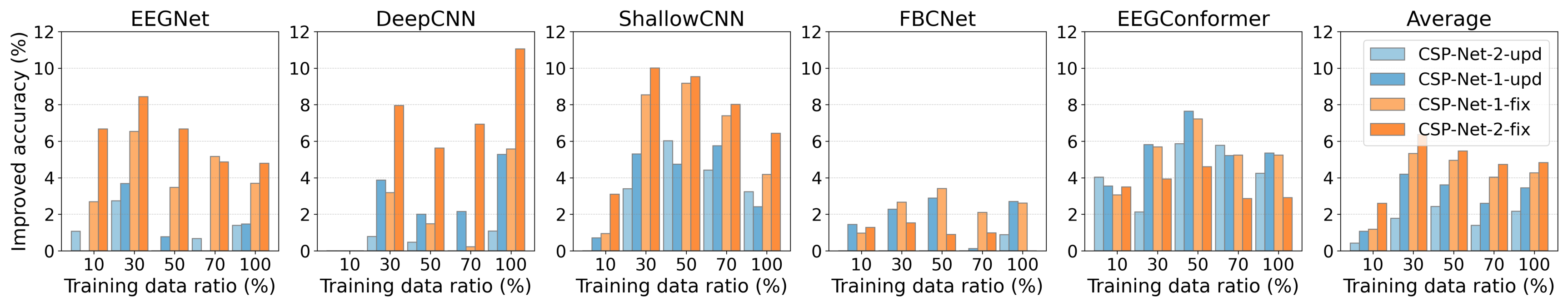}}
\subfigure[]{\label{fig:MI14S_cross}   \includegraphics[width=\linewidth,clip]{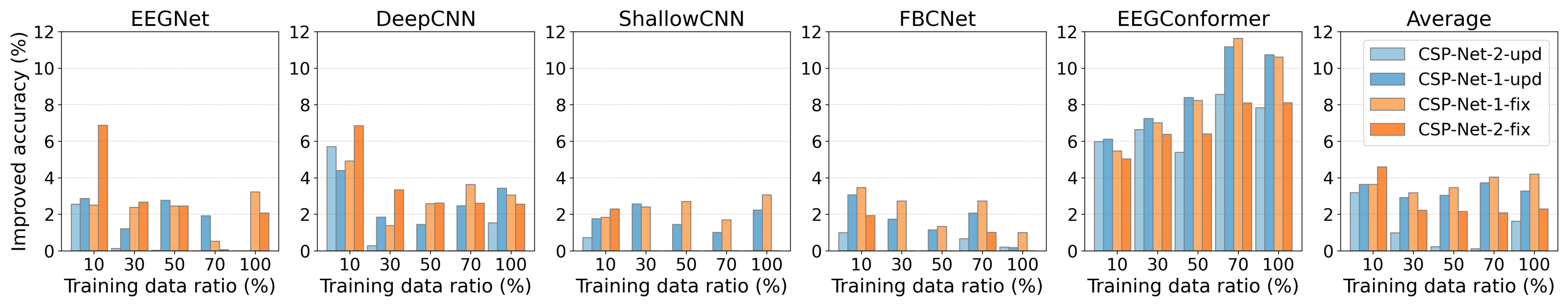}}
\caption{Accuracy improvements of CSP-Nets at different training data ratios on MI14S. (a) within-subject classification; and, (b) cross-subject classification.} \label{fig:data_ratio_MI14S}
\end{figure*}

\begin{figure*}[htbp]\centering
\subfigure[]{\label{fig:MI9S_within}  \includegraphics[width=\linewidth,clip]{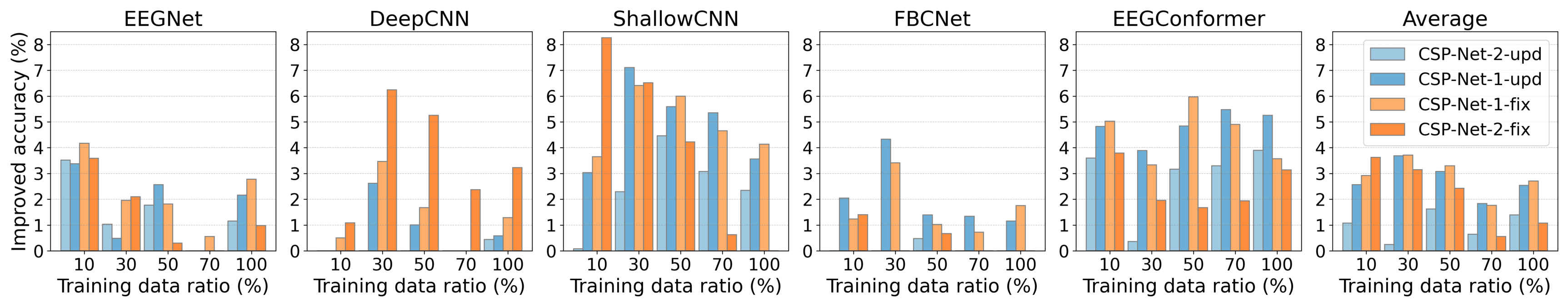}}
\subfigure[]{\label{fig:MI9S_cross}   \includegraphics[width=\linewidth,clip]{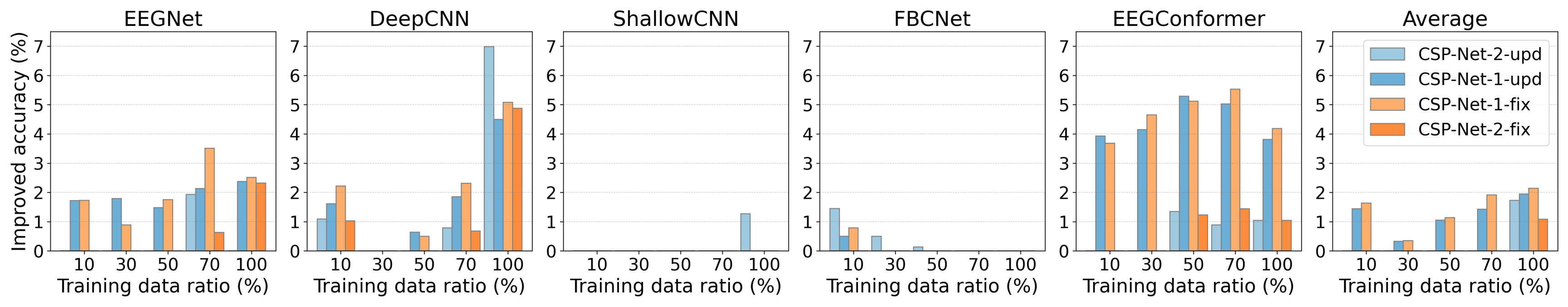}}
\caption{Accuracy improvements of CSP-Nets at different training data ratios on MI9S. (a) within-subject classification; and, (b) cross-subject classification.} \label{fig:data_ratio_MI9S}
\end{figure*}

Deep CNN models may easily overfit when the training dataset is small. Figs.~\ref{fig:data_ratio_MI4C}-\ref{fig:data_ratio_MI9S} show the accuracy improvements compared with the backbone at different training data ratios (the number of training samples used to train the model divided by the total number training samples) on the four datasets, respectively.

Observe that:
\begin{enumerate}
  \item Consistent with previous findings, CSP-Net-fix generally outperformed CSP-Net-upd, with CSP-Net-2-fix particularly competitive on EEGNet and DeepCNN. For example, in within-subject classification, CSP-Net-2-fix achieved a remarkable accuracy improvement of more than 20\% over the DeepCNN backbone, when trained with only 50\% of the training data on MI4C.
  \item Overall, the performance improvements of CSP-Nets were more obvious when the training data size was small. This may be because the embedding of prior knowledge greatly reduces the overfitting issue of CNN backbones in small sample scenarios.
\end{enumerate}

\subsection{Influence of the Number of CSP Filters} \label{sect:filter}

\begin{figure}[htpb]\centering
\subfigure[]{\label{fig:MI4C_within_filter}  \includegraphics[width=0.4\linewidth,clip]{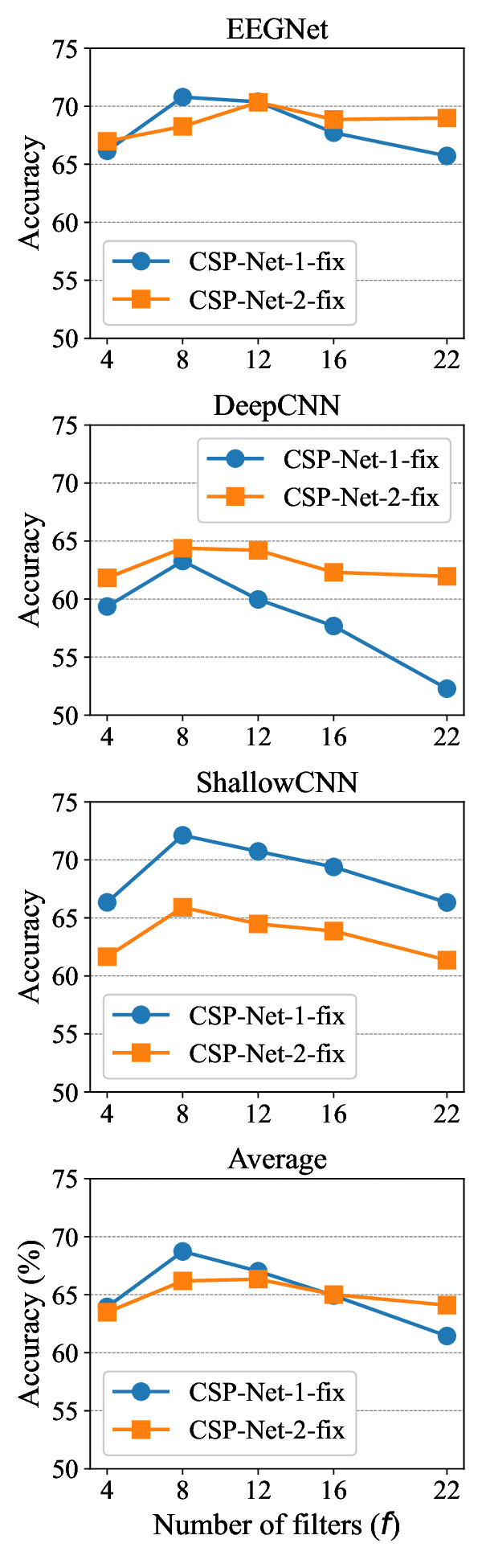}}
\subfigure[]{\label{fig:MI4C_cross_filter}   \includegraphics[width=0.4\linewidth,clip]{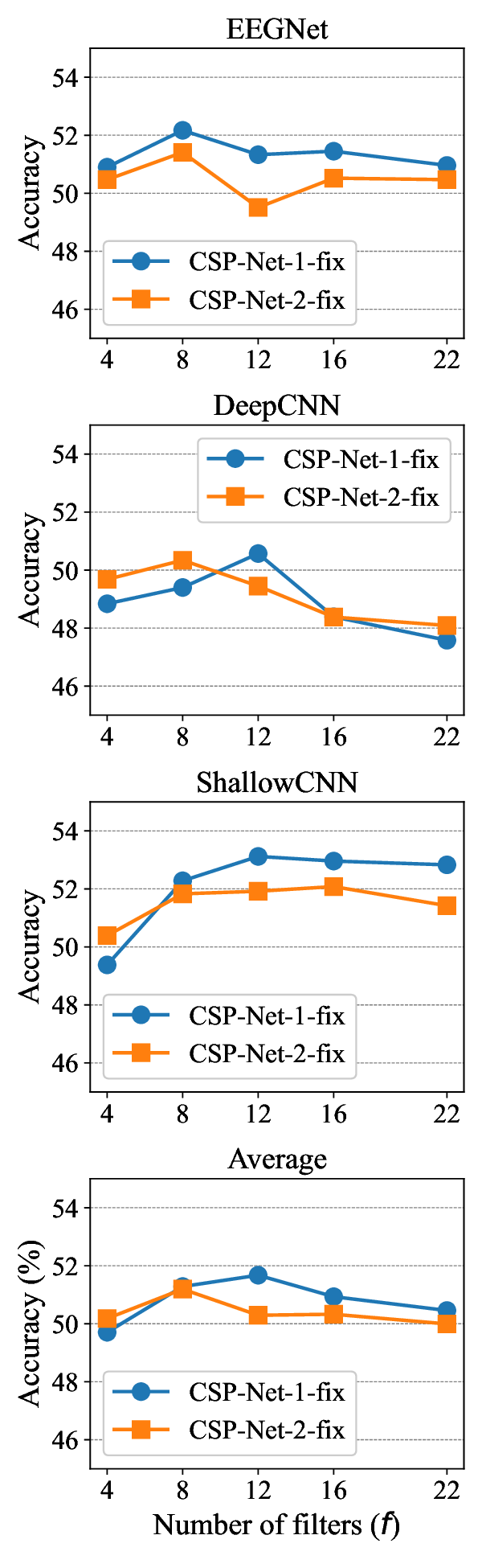}}
\caption{Classification accuracies of CSP-Nets using different number of CSP filters on MI4C. (a) within-subject classification; and, (b) cross-subject classification.} \label{fig:filter}
\end{figure}

We further investigated the influence of the number of CSP filters ($f$) on the performance of CSP-Nets with three backbones on MI4C. The dataset includes 22-channel EEG signals, so we considered $f\in\{4, 8, 12, 16, 22\}$. Fig.~\ref{fig:filter} shows the corresponding accuracies of the two CSP-Nets (fixed CSP layer). As the number of filters increased, the accuracy first increased and then decreases, which is intuitive. Generally, $f=8$ seems to be a good choice to balance the performance and computational cost.

\subsection{Ablation Studies}

An ablation study was performed to verify that the performance improvement of CSP-Net-1 was not due to an increase in the number of network parameters.

Specifically, we trained CSP-Net-1-rad, which replaced the CSP layer of CSP-Net-1 with a randomly initialized layer of the same size. The results on the four MI datasets in within-subject classification are shown in Table~\ref{tab:ablation_MI4C}. Generally, CSP-Net-1-rad performed similarly to the standard backbone, suggesting that the performance improvement of CSP-Net-1 was due to its incorporation of knowledge from CSP, instead of more parameters.

\begin{table}[htpb] \footnotesize
  \centering \setlength{\tabcolsep}{1mm} \renewcommand{\arraystretch}{0.9}
  \caption{Ablation study results of CSP-Net-1. Average accuracies higher than Standard are marked in bold.}
    \begin{tabular}{c|c|c|ccccc}
    \toprule
    \multirow{2}{*}{Dataset} & \multirow{2}{*}{Backbone} & \multirow{2}{*}{Approach} & \multicolumn{5}{c}{Training data ratio (\%)} \\ \cline{4-8}
    &   &   & 10   & 30   & 50   & 70   & 100  \\ \midrule
    \multirow{12}[2]{*}{MI4C} & \multirow{3}[2]{*}{EEGNet} & Standard & 37.67  & 45.60  & 54.29  & 57.61  & 63.50  \\
    & & CSP-Net-1-rad & 37.60  & 47.19  & 54.37  & 57.62  & 65.60  \\
    & & CSP-Net-1-fix & 43.39  & 57.75  & 62.28  & 65.72  & 70.79  \\ \cmidrule{2-8}
    & \multirow{3}[2]{*}{DeepCNN} & Standard & 27.17  & 28.90  & 32.13  & 44.66  & 51.51  \\
    & & CSP-Net-1-rad & 27.33  & 31.40  & 32.84  & 47.53  & 53.51  \\
    & & CSP-Net-1-fix & 30.97  & 36.95  & 43.56  & 56.09  & 63.28  \\ \cmidrule{2-8}
    & \multirow{3}[2]{*}{ShallowCNN} & Standard & 34.98  & 48.69  & 55.42  & 59.01  & 65.13  \\
    & & CSP-Net-1-rad & 36.78  & 47.59  & 54.16  & 60.68  & 64.69  \\
    & & CSP-Net-1-fix & 40.73  & 55.44  & 60.00  & 66.47  & 70.42  \\ \cmidrule{2-8}
    & \multirow{3}[2]{*}{Average} & Standard & 33.27  & 41.06  & 47.28  & 53.76  & 60.05  \\
    & & CSP-Net-1-rad & 33.90  & 42.06  & 47.12  & 55.28  & 61.27  \\
    & & CSP-Net-1-fix & \textbf{38.36}  & \textbf{50.05}  & \textbf{55.28}  & \textbf{62.76}  & \textbf{68.16}  \\ \midrule
    \multirow{12}[2]{*}{MI2C} & \multirow{3}[2]{*}{EEGNet} & Standard & 56.77  & 67.53  & 69.37  & 70.61  & 76.38  \\
    & & CSP-Net-1-rad & 57.77  & 66.76  & 68.51  & 71.85  & 77.11  \\
    & & CSP-Net-1-fix & 62.12  & 71.73  & 75.29  & 78.45  & 81.69  \\ \cmidrule{2-8}
    & \multirow{3}[2]{*}{DeepCNN} & Standard & 51.04  & 50.78  & 51.74  & 55.13  & 61.46  \\
    & & CSP-Net-1-rad & 51.97  & 52.50  & 54.14  & 56.94  & 62.69  \\
    & & CSP-Net-1-fix & 52.07  & 60.04  & 63.82  & 68.17  & 70.37  \\ \cmidrule{2-8}
    & \multirow{3}[2]{*}{ShallowCNN} & Standard & 51.38  & 62.34  & 68.36  & 71.70  & 76.29  \\
    & & CSP-Net-1-rad & 54.24  & 64.55  & 69.45  & 71.21  & 77.44  \\
    & & CSP-Net-1-fix & 62.42  & 72.12  & 76.69  & 81.37  & 83.71  \\ \cmidrule{2-8}
    & \multirow{3}[1]{*}{Average} & Standard & 53.06  & 60.22  & 63.16  & 65.81  & 71.38  \\
    & & CSP-Net-1-rad & 54.66  & 61.27  & 64.03  & 66.67  & 72.41  \\
    & & CSP-Net-1-fix & \textbf{58.87}  & \textbf{67.96}  & \textbf{71.93}  & \textbf{76.00}  & \textbf{78.59}  \\ \midrule
    \multirow{12}[2]{*}{MI14S} & \multirow{3}[2]{*}{EEGNet} & Standard & 59.10  & 66.99  & 72.25  & 73.59  & 75.22  \\
    & & CSP-Net-1-rad & 57.55  & 68.00  & 71.30  & 71.79  & 74.40  \\
    & & CSP-Net-1-fix & 61.79  & 73.53  & 75.72  & 78.76  & 78.92  \\ \cmidrule{2-8}
    & \multirow{3}[2]{*}{DeepCNN} & Standard & 51.11  & 50.80  & 53.92  & 56.64  & 55.75  \\
    & & CSP-Net-1-rad & 49.64  & 53.58  & 55.95  & 57.04  & 55.99  \\
    & & CSP-Net-1-fix & 49.33  & 53.99  & 55.40  & 56.87  & 61.33  \\ \cmidrule{2-8}
    & \multirow{3}[2]{*}{ShallowCNN} & Standard & 53.05  & 57.86  & 62.07  & 65.42  & 70.70  \\
    & & CSP-Net-1-rad & 53.08  & 57.48  & 61.79  & 66.19  & 71.56  \\
    & & CSP-Net-1-fix & 54.00  & 66.40  & 71.24  & 72.82  & 74.89  \\ \cmidrule{2-8}
    & \multirow{3}[1]{*}{Average} & Standard & 54.42  & 58.55  & 62.75  & 65.22  & 67.22  \\
    & & CSP-Net-1-rad & 53.42  & 59.69  & 63.01  & 65.01  & 67.32  \\
    & & CSP-Net-1-fix & \textbf{55.04}  & \textbf{64.64}  & \textbf{67.45}  & \textbf{69.48}  & \textbf{71.71}  \\ \midrule
    \multirow{12}[2]{*}{MI9S} & \multirow{3}[2]{*}{EEGNet} & Standard & 57.47  & 65.43  & 68.11  & 70.85  & 70.85  \\
    & & CSP-Net-1-rad & 59.03  & 64.89  & 67.19  & 69.97  & 69.57  \\
    & & CSP-Net-1-fix & 61.64  & 67.39  & 69.93  & 71.41  & 73.63  \\ \cmidrule{2-8}
    & \multirow{3}[2]{*}{DeepCNN} & Standard & 52.06  & 52.53  & 54.28  & 59.09  & 62.34  \\
    & & CSP-Net-1-rad & 51.93  & 53.01  & 54.22  & 56.45  & 61.92  \\
    & & CSP-Net-1-fix & 52.56  & 56.00  & 55.96  & 57.06  & 63.63  \\ \cmidrule{2-8}
    & \multirow{3}[2]{*}{ShallowCNN} & Standard & 51.89  & 57.55  & 61.90  & 66.04  & 69.03  \\
    & & CSP-Net-1-rad & 52.45  & 57.89  & 62.16  & 64.19  & 69.28  \\
    & & CSP-Net-1-fix & 55.54  & 63.96  & 67.90  & 70.70  & 73.17  \\ \cmidrule{2-8}
    & \multirow{3}[1]{*}{Average} & Standard & 53.81  & 58.50  & 61.43  & 65.33  & 67.41  \\
    & & CSP-Net-1-rad & 54.47  & 58.60  & 61.19  & 63.54  & 66.92  \\
    & & CSP-Net-1-fix & \textbf{56.58}  & \textbf{62.45}  & \textbf{64.60}  & \textbf{66.39}  & \textbf{70.14}  \\
    \bottomrule
    \end{tabular}%
  \label{tab:ablation_MI4C}%
\end{table}%

\subsection{Training Process Visualization}

\begin{figure*}[htpb]\centering
\subfigure[]{\includegraphics[width=\linewidth,clip]{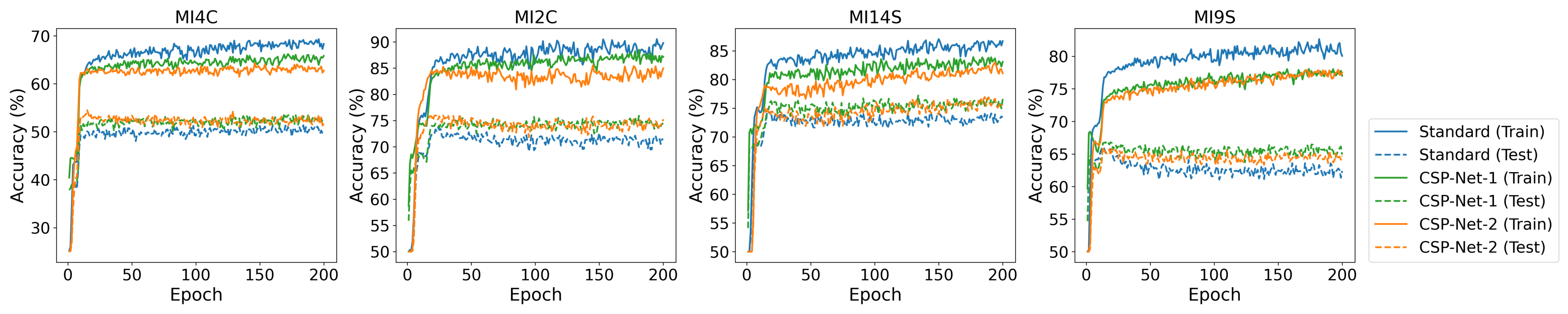}}
\caption{The training and test curves of different models on (a) MI4C; (b) MI2C; (c) MI14S; and, (d) MI9S.} \label{fig:curves}
\end{figure*}

Fig.~\ref{fig:curves} shows the average cross-subject training and test accuracy curves of CSP-Nets (fixed CSP layer) and their corresponding backbones (EEGNet) on the four datasets. For all the backbones, there was a large gap between the training and test curves, indicating overfitting. Our proposed CSP-Nets effectively leveraged the knowledge from the CSP filters for better initialization, reducing the gap and achieving better test performance.

\subsection{Visualization of the CSP Filters}

\begin{figure*}[htbp]\centering
\subfigure[]{\label{fig:MI2C_within_topo_cnn}  \includegraphics[width=\linewidth,height=0.09\textwidth,clip]{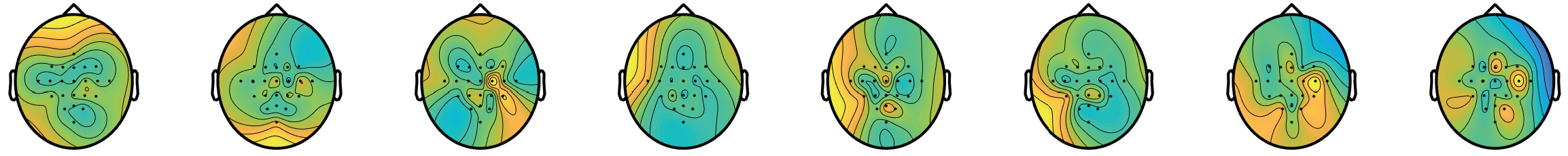}}
\subfigure[]{\label{fig:MI2C_within_topo_csp}  \includegraphics[width=\linewidth,height=0.09\textwidth,clip]{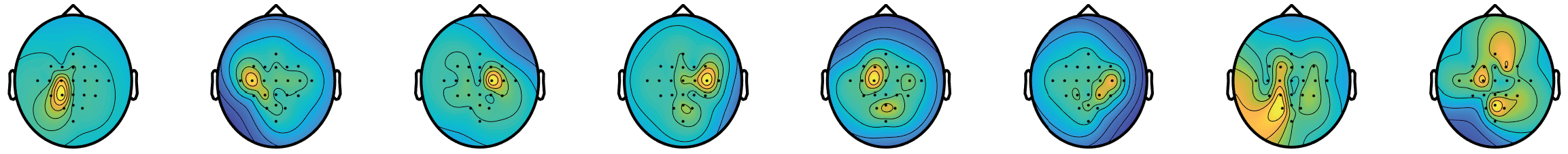}}
\caption{Visualization of eight (a) spatial filters in EEGNet and (b) the CSP filters in CSP-Net-2-fix for the within-subject classification on the Subject 1 of MI2C.} \label{fig:topoplot}
\end{figure*}

We further visualized the spatial convolutional kernel weights from the CSP-Net-2 and the counterparts from standard backbone. In Fig.~\ref{fig:topoplot}, we present the eight spatial filters in the EEGNet model and the CSP filters in the CSP-Net-2-fix model for within-subject classification on Subject 1 of MI2C (binary classification on the left hand and right hand). We can observe that the CSP filters in CSP-Net-2-fix exhibited a more focused and obvious left-right distribution concentrated on a specific sensorimotor area, which aligned well with the spatial characteristics of MI. In contrast, the standard EEGNet struggled to learn effective spatial information due to the absence of prior knowledge provided by CSP. This alignment suggests that CSP filters effectively capture the relevant spatial patterns inherent in the EEG data, enhancing the interpretability of the model.

\section{Conclusions} \label{sect:conclusion}

Spatial information, which can be well captured by CSP filters, is critical in EEG-based MI classification. This paper has introduced two CSP-Nets, which integrate the knowledge-driven CSP filters with data-driven CNN models. CSP-Net-1 directly adds a CSP layer before a CNN, utilizing CSP-filtered signals as input to enhance the discriminability. CSP-Net-2 replaces a convolutional layer in CNN with a CSP layer. Experiments on four public MI datasets demonstrated that the two CSP-Nets consistently improved over their CNN backbones, in both within-subject and cross-subject classifications. They are particularly useful when the number of training samples is very small. Remarkably, CSP-Net-1-fix, whose CSP layer uses fixed weights calculated using the traditional CSP algorithm, is the simplest yet demonstrates overall best performance.

Our work demonstrates the advantage of integrating knowledge-driven CSP filters with data-driven CNNs, or traditional machine learning with deep learning, in EEG-based BCIs.

\end{document}